\documentclass[aps,twocolumn,showpacs]{revtex4}

\usepackage{graphicx}%
\usepackage{dcolumn,float}
\usepackage{amsmath,amssymb}

\begin{document}

\title{Stripe Ansatzs from Exactly Solved Models
}

\date{\today}

\author{  
M.A. Mart\'{\i}n-Delgado$^{\ast1}$, M. Roncaglia$^{\ast2}$ and
G. Sierra$^{\ast3}$
 } 
\affiliation{ 
$^{\ast1}$Departamento de
F\'{\i}sica Te\'orica I, Universidad Complutense,  Madrid, Spain.
\\ 
$^{\ast2}$Dipartimento di Fisica, Universit\`a di Bologna, INFN and
INFM, Bologna, Italy.\\
$^{\ast3}$Instituto de Matem\'aticas y F\'{\i}sica Fundamental, C.S.I.C.,
Madrid, Spain. }

\begin{abstract}
Using the Boltzmann weights of classical Statistical
Mechanics vertex models we define a new class of
Tensor Product Ansatzs for 2D quantum lattice systems, 
characterized by a 
strong anisotropy, which gives rise to stripe
like structures. In the case of the six vertex model
we compute exactly, in the thermodynamic limit,
the norm of the ansatz and other observables.
Employing this ansatz we study the phase diagram 
of a Hamiltonian given by the sum of XXZ 
Hamiltonians along the legs coupled by an Ising term.
Finally, we suggest a connection between
the six and eight-vertex Anisotropic Tensor Product
Ansatzs, and their associated Hamiltonians, with the
smectic stripe phases recently discussed in the literature. 
\end{abstract}


\maketitle

\newcommand{\ba}{\boldsymbol{\alpha}}
\newcommand{\bb}{\boldsymbol{\beta}}
\newcommand{\be}{\boldsymbol{\eta}}
\newcommand{\bx}{\boldsymbol{\xi}}
\newcommand{\0}{\boldsymbol{0}}
\newcommand{\1}{\boldsymbol{1}}
\newcommand{\T}{T^{\bf Q}}

\section{Introduction}
\label{sec1:level1}

Low dimensional spin systems constitute one of the most
active areas in Condensed Matter Physics due to the experimental
findings and the associated theoretical activity.
These systems are strongly correlated with
very rich phase diagrams studied by means
of a miscellanea of analytical and numerical
techniques,   among which the study of simplified
variational ansatzs for the ground state (GS) and excitations
have played a significant role. 

In 1D there is a plethora
of variational ansatzs: the AKLT states \cite{aklt}, the finitely
correlated ansatzs \cite{fannes}, 
the Matrix Product Ansatzs (MPA) \cite{klumper,ostlund,roman}, 
the Recurrent Variational Ansatz (RVA) \cite{rva}, etc. 
All these ansatzs have a common structure for 
the GS wave function  which is given by  the sum,
over some auxiliary variables, 
of products of amplitudes that also depend on the 
spin variables at the sites. The basic quantity
here is the ``matrix product 
amplitude''  $A_{\alpha, \beta}[m_i]$ 
where $m_i$ is the spin at the $i^{\rm th}$  site and $\alpha$
and $\beta$ are auxiliary variables, which can be
associated to the links meeting  at the site.  
For a spin chain
with $N$ sites and
periodic boundary conditions the corresponding
state can be written as \cite{ostlund}

\begin{equation}
|\psi_{\rm MPA} \rangle  = \sum_{m_i's} 
{\rm Tr}( A[m_1] \dots A[m_N]) |m_1 \rangle 
\dots |m_N \rangle 
\label{1}
\end{equation}

\noindent where the trace is over the auxiliary variables $\alpha$.
Some MPA states, such as the AKLT ones,  
are exact ground states of a Hamiltonian, which is 
given by the sum of projectors between nearest neighbours sites \cite{aklt}. 
In other cases the MPA states are used as variational ansatzs
for a given Hamiltonian,  with the MPA amplitudes
$A_{\alpha, \beta}[m]$ playing the role of variational parameters.
Within the latter category fall the DMRG states \cite{white} 
(for a review on the DMRG see \cite{dresden,karen}),  
which are in fact MPA states with open 
boundary conditions and position dependent 
amplitudes ( i.e. inhomogenous MPA's) 
\cite{ostlund,dukelsky,tasaki}. In the DMRG 
the auxiliary variables label the states kept in the blocks.

The MPA states can be generalized in a natural way
to 2D systems, replacing the matrix amplitudes
$A_{\alpha, \beta}[m]$ by ``tensor product'' amplitudes
$A_{\alpha_1, \alpha_2, \dots, \alpha_z}[m]$, 
where $z$ is given 
by the coordination  number of the lattice, i.e.
$z=3$ for an hexagonal lattice, $z=4$ for  
a square lattice and so on so forth 
\cite{niggemann97,nishino98,niggemann98,
faro,dziurzik99,hieida,nishino00a,nishino00b,vdma}. 
By analogy
with Statistical Mechanics (SM) these states can be called
Tensor Product Vertex Ansatzs (TPVA) because
the auxiliary variables $\alpha_i$ are associated
to the links of the lattice, while  the amplitudes
are associated to the vertices \cite{baxter}. Another class 
of 2D ansatzs is formed 
by the Tensor Product Face Ansatzs (TPFA),  where
the amplitudes are associated to the faces of a square lattice
as in the Face or Interaction Around a Face (IRF) 
models in Statistical Mechanics \cite{faro,okunishi99}. 
The 2D generalizations of the AKLT states for spins
3/2, 2 and higher belong to the TPVA class. 
The recipy to construct TPA's is as
in equation (\ref{1}), where  the contraction of the auxiliary
variables follows the  pattern 
of the underlying vertex or face models. 

Most of the TPA's studied in the literature
are isotropic, meaning that their properties
are largely independent of the spatial direction. 
However in 2D and 3D there are physical systems,
like some high-temperature superconductors \cite{highTc},  
quantum Hall systems \cite{hall}, or manganites \cite{manganites},  
which exhibit strongly anisotropic
properties due to the existence of stripes. 
These objects are static or dynamic charge inhomogeneities, 
which are linear in 2D or planar in 3D. 
One may wonder whether these systems can be modelled
with simple TPA's, just as Haldane spin chains can
be easily described as valence-bond states. 
In this work we shall not address directly this question,   
but the results we have obtained suggest the possibility of 
a simple description of stripes in terms
of TPA's. 
More precisely, in this paper we shall investigate 
a class of TPA's 
based on classical exactly solvable 2D vertex models
with strong anisotropic properties reminiscent
to the stripe systems investigated in reference 
\cite{emery00,carpentier}. 
Any classical SM
2D vertex model, not necessarily integrable,  
defined by its Boltzmann weights, 
give rise to an Anisotopic Tensor Product 
Ansatz (ATPA). If, in addition, the SM model is exactly
solvable,  then the corresponding ATPA becomes   
quasi exactly solvable. 
The latter term is borrowed from the theory
of spectral problems associated to the Schr\"odinger equation
\cite{qes} 
meaning, in our context, that
some quantities, as the norm of the ATPA's 
and some expectation values, can be computed
exactly in the thermodynamic limit.

To illustrate our proposal   we have choosen 
the well known 6 vertex model,  whose 1D quantum
mechanical counterpart is the XXZ model or
the 1D spinless fermion \cite{baxter}. We shall show that the
corresponding ATPA  has some similarities with
the striped states of 2D spinless fermions 
studied in the literature \cite{emery00,carpentier}.

The organization of this paper is as follows.
First of all we review briefly the basic ingredients
of vertex models in Statistical Mechanics
( section II) 
and the tensor product vertex ansatzs 
( section III). In section IV we introduce 
the ATPA's based on SM vertex models
and study their general properties. The ATPA
associated to the six-vertex model
is used  in section V as a trial ground state
for an anisotropic Hamiltonian 
closely related to the XXZ spin chain Hamiltonian, 
and derive the phase diagram. 
In section VI we briefly comment on the
eigh-vertex ATPA model. The possible
connections between the six and eight-vertex
ATPA is explored in section VII and finally 
in section VIII
we state our conclusions. In Appendices A
and B  we collect  some technical results.

\section{Vertex Models in Statistical Mechanics}
\label{sec2:level1}

Throughout this paper we shall follow closely Baxter's book
\cite{baxter}. 
Let us consider a rectangular lattice with $N$ rows 
and $L$ columns. Throughout these paper we shall also
use the term ``legs'' for the rows and ``rungs'' for the
columns.  In a vertex model there is a local state 
variable $\alpha$  associated to every  link and a Boltzmann 
weight associated to every vertex $\bullet$,  
which depends on the 
four link variables meeting at it. We shall represent 
the Boltzmann weight as

\begin{equation}
 W_{\alpha, \xi}^{\beta,  \eta} = \begin{array}{ccccc}
& &  \eta &  & \\
& &  | &  & \\
\alpha & - & \bullet &    - & \beta \\
& &  | &  &  \\
& &   \xi &   & \\
\end{array}
\label{2}
\end{equation}

The statistical  weight of a global configuration is given
by the product of the Boltzmann weights of all the vertices.
The partition function $Z$ is the sum of these weights 
over all the link configurations, which
can also be expressed  using transfer matrices.
The row-to-row and column-to-column transfer matrices
are defined as,

\begin{align}
& T^{\rm row}_{\be,\bx} = 
\sum_{\alpha's} \; \prod_{i=1}^{L}  
W_{\alpha_i, \xi_i}^{\alpha_{i+1},  \eta_i} 
\label{3} \\
& T^{\rm col}_{\ba,\bb} = 
\sum_{\xi's} \; \prod_{i=1}^{N}  
W_{\alpha_i,  \xi_i }^{ \beta_{i},  \xi_{i+1}} 
\nonumber 
\end{align}

\noindent  where $\bx = (\xi_1, \dots, \xi_L)$, 
$\ba = (\alpha_1, \dots, \alpha_N)$, etc and 
the  periodic boundary conditions are assumed
along  both directions. Using (\ref{3}) 
the partition function $Z$ reads,

\begin{equation}
Z = {\rm Tr} \; T_{\rm row}^N =  {\rm Tr} \; T_{\rm col}^L
\label{4}
\end{equation}

As an example we display  in table 1
the Boltzmann weights for the allowed vertex
configurations  
of the six-vertex model. 
The link variables take on two values, 
say 0 and 1, 
which in the standard notation correspond to 
the right and up pointing ( for  $\alpha=0$),  
and left and down pointing (for $\alpha=1$).  
The allowed configurations 
satisfy the ice rule $\alpha + \xi = 
\beta + \eta $,  and the Boltzmann weights
are invariant under the reversal of all arrows, which
leaves three independent ones, called  $a,b$ and $c$. 
The 6-vertex  model is integrable: there is a 
uniparametric family of transfer matrices $T(u)$ 
commuting among themselves. 
This is guaranteed by the Yang-Baxter equation
satisfied by the Boltzmann weights.

\begin{center}
\begin{tabular}{|c|c|c|c|}
\hline
weight &  $a$ & $b$ & $c$  \\
\hline
 $\begin{array}{ccc}
& \eta  &  \\
\alpha  &  & \beta   \\
& \xi  &   \\
\end{array}$ 
& $\begin{array}{ccc}
& 0  &  \\
0  &  & 0   \\
& 0  &   \\
\end{array}$ & 
$\begin{array}{ccc}
& 0  &  \\
1  &  & 1   \\
& 0  &   \\
\end{array}$ & 
$\begin{array}{ccc}
& 0  &  \\
0  &  & 1   \\
& 1  &   \\
\end{array}$  \\ \cline{2-4} 
& $\begin{array}{ccc}
& 1  &  \\
1  &  & 1   \\
& 1  &   \\
\end{array}$ & 
$\begin{array}{ccc}
& 1  &  \\
0  &  & 0   \\
& 1  &   \\
\end{array}$ & 
$\begin{array}{ccc}
& 1  &  \\
1  &  & 0   \\
& 0  &   \\
\end{array}$ \\ \hline 
\end{tabular}

Table 1. 
Boltzmann weights of the six-vertex model 
\end{center}

\section{Tensor Product Vertex Ansatzs}
\label{sec3:level1}

As in the previous section we shall consider
a lattice with $N$ legs of length $L$. In the
quantum spin model there is a spin degree
of freedom  $m$ 
at each vertex of the lattice. 
To construct a TPA we shall
associate an auxiliary variable
$\alpha$ to each link, as in the SM models. 
The TPA amplitudes will be denoted as 

\begin{equation}
A_{\alpha \xi}^{\beta  \eta}[m] = \begin{array}{ccccc}
& & \eta & & \\
& & | & & \\
\alpha & - & m & - & \beta \\
& & | & &  \\
& &  \xi & & \\
\end{array}
\label{5}
\end{equation}

By analogy with SM we shall define the 
row-to-row and column-to-colum transfer matrix
amplitudes

\begin{align}
& A^{\rm row}_{\be,\bx}[{\bf m}] = 
\sum_{\alpha's} \; \prod_{i=1}^{L}  
A_{\alpha_i, \xi_i}^{\alpha_{i+1},  \eta_i}[m_i] 
\label{6} \\
& A^{\rm col}_{\ba,\bb}[{\bf m}] = 
\sum_{\xi's} \; \prod_{i=1}^{N}  
A_{\alpha_i,  \xi_i }^{ \beta_{i},  \xi_{i+1}}[m_i] 
\nonumber 
\end{align}

\noindent where ${\bf m}= (m_1, \dots, m_L)$ for  $A^{\rm row}$
while  ${\bf m}= (m_1, \dots, m_N)$ for  $A^{\rm col}$. 
Using eqs.(\ref{6}) the TPA can be written 
in two alternative ways, i.e.

\begin{align}
& | \psi \rangle_{\rm row} = \sum_{ {\bf m}'s} {\rm Tr}
\left( A^{\rm row}[{\bf m_1}] \dots A^{\rm row}[{\bf m_N}] \right)
|{\bf m_1},  \dots,  {\bf m_N} \rangle_{\rm row}
\label{7} \\
& | \psi \rangle_{\rm col} = \sum_{ {\bf m}'s} {\rm Tr}
\left( A^{\rm col}[{\bf m_1}] \dots A^{\rm col}[{\bf m_L}] \right)
|{\bf m_1}, \dots , {\bf m_L} \rangle_{\rm col}
\nonumber 
\end{align}

\noindent 
These equations are identical to  eq.(\ref{1}), 
which implies that the  TPA can be regarded as a MPA 
through  the legs or the rungs.
The norm of the TPA  is given by,

\begin{equation}
\langle \psi|\psi \rangle = {\rm Tr} \; {\cal T}_{\rm row}^N = 
{\rm Tr} \; {\cal T}_{\rm col}^L
\label{8}
\end{equation}

\noindent where

\begin{align}
& {\cal T}^{\rm row}_{ \be \be', \bx  \bx'}=
\sum_{\bf m} \; 
A^{\rm row}_{\be,\bx}[{\bf m}] 
A^{\rm row}_{\be',\bx'}[{\bf m}] 
\label{9} \\
& {\cal T}^{\rm col}_{ \ba \ba', \bb  \bb'}=
\sum_{\bf m} \; 
A^{\rm col}_{\ba,\bb}[{\bf m}] 
A^{\rm col}_{\ba',\bb'}[{\bf m}] 
\nonumber
\end{align}

Thus the computation of the 
norm (\ref{8}) amounts to that  
of the partition function of a classical  SM  vertex model where
the link variables are twice those of the
quantum mechanical model.

\section{Anisotropic Tensor Product Ansatzs}
\label{sec4:level1}

\subsection{Generic Case}
\label{sec4:level2}

Let us suppose we are given a vertex model
with Boltzmann weights 
$ W_{\alpha, \xi}^{\beta,  \eta}$.
Using them, we shall define a
ATPA model by the equation:

\begin{equation}
A_{\alpha, \xi}^{\beta,  \eta}[m] = \delta_{m, \eta}
\; W_{\alpha, \xi}^{\beta,  \eta}
\label{10}
\end{equation}

\noindent where the spin variable
at each site, i.e. $m$, is identified with the 
link variable $\eta$. In the case of the
six-vertex model the shall adopt the convention that
$m=0$ corresponds to spin $1/2$  and $m=1$ to spin $-1/2$.
For the six-vertex model the corresponding TPA
amplitudes are given in table 2.

\begin{center}
\begin{tabular}{|c|c|c|c|}
\hline
Amplitude &  $a$ & $b$ & $c$  \\
\hline
 $\begin{array}{ccc}
& \eta  &  \\
\alpha  & m & \beta   \\
& \xi  &   \\
\end{array}$
 & $\begin{array}{ccc}
& 0  &  \\
0  & + & 0   \\
& 0  &   \\
\end{array}$ & 
$\begin{array}{ccc}
& 0  &  \\
1  & + & 1   \\
& 0  &   \\
\end{array}$ & 
$\begin{array}{ccc}
& 0  &  \\
0  &+  & 1   \\
& 1  &   \\
\end{array}$  \\ \cline{2-4} 
& $\begin{array}{ccc}
& 1  &  \\
1  &-  & 1   \\
& 1  &   \\
\end{array}$ & 
$\begin{array}{ccc}
& 1  &  \\
0  &-  & 0   \\
& 1  &   \\
\end{array}$ & 
$\begin{array}{ccc}
& 1  &  \\
1  &-  & 0   \\
& 0  &   \\
\end{array}$ 
 \\ \hline 
\end{tabular}

Table 2.
Six-vertex TPA amplitudes  
\end{center}

The choice (\ref{10}) is extremely anisotropic
since it treats on a  very different footing the vertical
and horizontal directions of the lattice. 
This is the main reason to consider both the
row-to-row and column-to-column transfer matrices, 
which give rice to complementary  descriptions of  the ansatz.
In SM models where the leg and rung variables 
run over different sets,   
one can obtain  two inequivalent ATPA's,  
not related by a 90º degrees rotation. 
In the rest of the paper we shall suppose that 
the link variables are of the same type 
in both directions. 

Eq.(\ref{10}) implies a simple relationship  
between the row-to-row TPA amplitude (\ref{6}) 
and the row-to-row transfer matrix  (\ref{3})
of the underlying SM model, namely

\begin{equation}
A^{\rm row}_{\be,\bx}[{\bf m}] = 
\delta_{{\bf m}, \be} \; 
 T^{\rm row}_{\be,\bx}
\label{11}
\end{equation}

\noindent which leads to the following row representation
of the ATPA state (\ref{7}),

\begin{equation}
 | \psi \rangle_{\rm row} = \sum_{ {\bf m}'s} 
 T^{\rm row}_{{\bf m_1}, {\bf m_2}}  
T^{\rm row}_{{\bf m_2}, {\bf m_3}} 
 \dots T^{\rm row}_{{\bf m_N}, {\bf m_1}} 
|{\bf m_1},  \dots,  {\bf m_N} \rangle_{\rm row}
\label{12}
\end{equation}

\noindent where  $T^{\rm row}_{{\bf m_1}, {\bf m_2}}$
is the row-to-row  transfer matrix
(\ref{3}) built up from the SM Boltzmann weights
$ W_{\alpha, \xi}^{\beta,  \eta}$
defining the ATPA ( eq.(\ref{10})). 
The structure of this state is similar  to the 
Kramers-Wannier variational state, first proposed
for the GS of the Ising model \cite{kramers,nishino98}, 
where the analogue
of  $T^{\rm row}$ is played by $2 \times 2$ matrices. 
The ATPA built from the choice (\ref{10}) can be
seen as a superposition of leg states connected through
the row-to-row transfer matrix of the SM model. For example, 
in the antiferroelectric phase of the six-vertex model, 
the state along the legs will be mostly of Neel type
and correlated antiferromagnetically with their nearest
neighbours legs. In the spinless fermion picture
the latter state is a Wigner crystal with CDW order. 

The norm of (\ref{12}) is given simply by

\begin{equation}
\langle \psi  | \psi \rangle_{\rm row} = \sum_{ {\bf m}'s} 
 (T^{\rm row}_{{\bf m_1}, {\bf m_2}})^2  
(T^{\rm row}_{{\bf m_2}, {\bf m_3}})^2 
 \dots (T^{\rm row}_{{\bf m_N}, {\bf m_1}})^2 
\label{13}
\end{equation}

\noindent 
It is important to notice that (\ref{13}) is not the 
partition function of the SM model defined with  
Boltzmann weights 
$ W_{\alpha, \xi}^{\beta,  \eta}$ or their square.
The reason being that in general,

\begin{equation}
 (T^{\rm row}_{{\bf m_1}, {\bf m_2}})^2
\neq 
 (T_{\rm row}^2)_{{\bf m_1}, {\bf m_2}} 
\label{14}
\end{equation}

\noindent where the LHS of this equation 
is the square of the
element  $ T^{\rm row}_{{\bf m_1}, {\bf m_2}}$ 
of the row-to-row transfer matrix,  while
the RHS is the entry $({\bf m_1}, {\bf m_2})$ 
of the square of the row-to-row transfer matrix. 
In any case, the computation of (\ref{13}) requires
much less effort than eqs.(\ref{8}) because the 
matrices involved contain half of the indices of those
of the general case. In other
words, the ATPA does not lead
to a doubling of indices in the row representation. 

The situation 
improves even further in the column representation.
Using eqs.(\ref{6}) and ({\ref{10}) we
see that the  
column ATPA amplitudes are given by the product
of the Boltzmann weights on a column, i.e.

\begin{equation}
 A^{\rm col}_{\ba,\bb}[{\bf m}] = 
 \prod_{i=1}^{N}  
 W_{\alpha_i, { m_{i-1}}}^{\beta_i,  { m_i}}
\label{15}
\end{equation}

\noindent where $m_0 = m_N$. Consequently the
column-to-column ATPA transfer matrix (\ref{9}) 
becomes

\begin{equation}
 {\cal T}^{\rm col}_{ \ba \ba', \bb  \bb'}=
\sum_{\bf m} \; \prod_{i=1}^N 
 W_{\alpha_i, { m_{i-1}}}^{\beta_i,  { m_i}} 
\;  W_{\alpha'_i, { m_{i-1}}}^{\beta'_i,  { m_i}}
\label{16}
\end{equation}

Let us suppose for a moment  that we restrict ourselves
to the ``diagonal'' sector of $ {\cal T}^{\rm col}$, which
is defined by the choices $\ba = \ba'$ and $\bb = \bb'$. Then 
(\ref{16}) becomes a  column-to-column 
transfer matrix (\ref{3}) with Boltmann weights being 
the square of $ W_{\alpha, \xi}^{\beta,  \eta}$, namely

\begin{equation}
 {\cal T}^{\rm col}_{ \ba \ba, \bb  \bb}=
\sum_{\bf m} \; \prod_{i=1}^N 
\left(  W_{\alpha_i, { m_{i-1}}}^{\beta_i,  { m_i}} \right)^2
=  T^{\rm col}_{ \ba , \bb}( W^2) 
\label{17}
\end{equation}

For a generic TPA  this diagonal truncation may be a good
approximation in certain regions of the parameter space, 
as has been shown by  Niggemann
et al.  in a TPA for a spin 3/2 system
on a hexagonal latice \cite{niggemann97}. 

We shall show below that for a subclass of ATPA's,
this diagonal truncation is in fact exact, which 
has important consequences. 

\subsection{Ansatzs with conserved quantum numbers}
\label{sec4:level3}

Let us  assume that 
the Boltmann weights 
$ W_{\alpha, \xi}^{\beta,  \eta}$ satisfy
a conservation law of the type,

\begin{equation}
W_{\alpha, \xi}^{\beta,  \eta} = 0
\;\; {\rm unless} \;\; \alpha + \xi = \beta + \eta 
\label{18}
\end{equation}

\noindent where the link variables label the basis
of an irreducible representation (irrep)  
of a Lie group ${\cal G}$.
The six-vertex model corresponds
to the  spin 1/2 irrep of the group ${\cal G}= SU(2)$,
with the convention  $\alpha =0$ (resp. 1)  for the
$s_z=1/2$ ( resp. $s_z=-1/2)$. 
For a general Lie group the link variables 
will be given by  
the weights of the corresponding irrep.

The immediate  consequence of (\ref{18}) is that 
the non vanishing terms of (\ref{16}) must satisfy

\begin{align}
& \alpha_i + m_{i-1} = \beta_i + m_i ,\;\; i = 1, \dots, N
\label{19} \\
& \alpha'_i + m_{i-1} = \beta'_i + m_i 
\nonumber 
\end{align}

\noindent which implies 

\begin{equation}
\alpha'_i - \alpha_i = \beta'_i - \beta_i = Q_i 
\label{20}
\end{equation}

\noindent where $Q_i=1,0, -1$ for the six-vertex model. 
In the general case  
$Q_i$, being the difference of two weights of irreps, 
is either zero or a root of the Lie group ${\cal G}$.

Defining the Boltzmann weights $W^Q$ as

\begin{equation}
{(W^Q)}_{\alpha, \xi}^{\beta,  \eta} = 
W_{\alpha, \xi}^{\beta,  \eta} \; 
W_{\alpha + Q, \xi}^{\beta + Q,  \eta}
\label{21}
\end{equation}

\noindent we see from (\ref{16}) and (\ref{20})
that  ${\cal T}^{\rm col}$ breaks into
block transfer matrices $\T$  labelled by the vector
${\bf Q} = (Q_1, \dots, Q_N)$, whose entries are given by

\begin{equation}
 \T_{ \ba , \bb } \equiv 
 {\cal T}^{\rm col}_{ \ba \ba + {\bf Q}, \bb  \bb + {\bf Q}}=
\sum_{\bf m} \; \prod_{i=1}^N 
 (W^{Q_i})_{\alpha_i, { m_{i-1}}}^{\beta_i,  { m_i}} 
\label{22}
\end{equation}

The case ${\bf Q} = \0$ corresponds to the 
matrix (\ref{17}), and hence the truncation of
the model to the ``diagonal'' sector is not an approximation, 
but an exact result. This fact greatly simplifies the  
computation of the  norm  of the ATPA 
in the thermodynamic limit $L \rightarrow \infty$, 
which is given by $\Lambda_{\rm max}^L $ where $\Lambda_{\rm max}$
is  the biggest of all   
largest eigenvalues $\Lambda^{\bf Q}_0$ of  the matrices $\T$. 
In appendix A we show that this eigenvalue belongs to the ${\bf Q} = 0$ sector
and thus,

\begin{equation}
\lim_{L \rightarrow \infty} 
\langle \psi | \psi \rangle_{\rm col} 
= \Lambda_{\rm max}^L, \;\;  \Lambda_{\rm max}= \Lambda^{\bf Q=0}_0 
\label{23}
\end{equation}

This is quite a useful result for it implies that if the SM model
defined by the Boltzmann weights $W^0= (W)^2$ (\ref{21}) is integrable
then $\Lambda_{\rm max}$ can be 
computed exactly, at least in the limit $N\rightarrow \infty$. 
In the case
of the ATPA based on the six-vertex model, the Boltzmann weights
$W^0$ are simply the square of the original ones, i.e.

\begin{equation}
 W^0(a,b,c) = W(a^2,b^2,c^2) 
\label{24}
\end{equation}

\noindent and hence the norm of the ATPA 
can be computed exactly in the thermodynamic limit.

There is yet another important consequence of the conservation
law (\ref{18}). As it is well known in the theory of transfer matrices,
eq.(\ref{18}) implies that the row-to-row transfer matrix
preserves the sum of all quantum numbers of every row, i.e.

\begin{equation}
T^{\rm row}_{ {\bf m}, {\bf m'}} = 0 \;\; {\rm unless}
\;\; \sum_{i=1}^L m_i =  \sum_{i=1}^L m'_i
\label{25}
\end{equation}

\noindent 
Hence all the terms appearing in the sum
(\ref{12}), giving $|\psi \rangle_{\rm row}$,  
must have the same value of ``angular momenta'' per leg.  
In the six-vertex model this implies the vanishing of all 
correlators between raising and lowering spin 
operators among different rows/legs, i.e.

\begin{equation}
\langle S^+_{i,a} \;  S^-_{j,b} \rangle = 0 , \;\; {\rm if} \;\; i \neq j
\label{26}
\end{equation}

\noindent where $S^\pm_{i,a}$ is the raising (lowering)
spin operator on the $a^{\rm th}$  site of the 
$i^{\rm th}$ leg. In other words, the  quantum
fluctuations across the legs of the ATPA are strictly forbidden.
In the six vertex model the spins may only fluctuate along the legs. 
Using the spinless fermion terminology, the only allowed
charge fluctuations occur inside the legs. As  mentioned
in the introduction, this lack of quantum fluctuations
across the legs is reminiscent to  that occurring 
in some models of high-$T_c$ superconductors ( see section 
VII).

The previous considerations give us a hint on what sort
of Hamiltonians the ATPA's may become approximate ground
states. After all, we want to use the ATPA  as
variational ansatzs for physically interesting systems. 
We postpone this question until next section 
after a discussion on correlators
and density matrices for ATPA's.

\subsection{Correlators and density matrices}
\label{sec4:level4}

Let ${\cal O}_{i,i+1}^{\rm d}$ be a diagonal 
operator acting between the legs $i$ and $i+1$ that
do not change their states and with matrix element 
$ \hat{{\cal O}}_{{\bf m_i},{\bf m_{i+1}}}^{\rm d}$. 
A typical example in the six-vertex model
is provided by the Ising term $\sigma^z_{i,a} \sigma^z_{i+1,a}$.
The expectation value of ${\cal O}_{i,i+1}^{\rm d}$  
in the ATPA is be given by,

\begin{equation}
\begin{aligned}
\langle \psi|  {\cal O}_{i,i+1}^{\rm d}  
| \psi \rangle_{\rm row} &  = 
 \sum_{ {\bf m}'s} 
 (T^{\rm row}_{{\bf m_1}, {\bf m_2}})^2  
(T^{\rm row}_{{\bf m_2}, {\bf m_3}})^2 \\
&  \dots (T^{\rm row}_{{\bf m_N}, {\bf m_1}})^2 
\hat{{\cal O}}_{{\bf m_i},{\bf m_{i+1}}}^{\rm d}
\end{aligned}
\label{27}
\end{equation}

It can be shown that the square
of the row-to-row  transfer matrix can be written
as ( recall eqs.(\ref{20}) and (\ref{21})) 

\begin{equation}
 (T^{\rm row}_{{\bf m_1}, {\bf m_2}})^2
= \sum_Q 
 T(W^Q)_{{\bf m_1}, {\bf m_2}}  
\label{28}
\end{equation}

\noindent and hence the sum (\ref{27}) becomes

\begin{equation}
\begin{aligned}
\langle \psi|  {\cal O}_{i,i+1}^{\rm d}  
| \psi \rangle_{\rm row} & = 
\sum_{ {\bf m}'s} \sum_{Q_1, \dots, Q_N}
T(W^{Q_1})_{{\bf m_1}, {\bf m_2}} \\
 &\dots T(W^{Q_N})_{{\bf m_N}, {\bf m_1}} 
\hat{{\cal O}}_{{\bf m_i},{\bf m_{i+1}}}^{\rm d}
\end{aligned}
\label{29}
\end{equation}

\noindent 
In the thermodynamic limit this sum will 
be dominated by the term $Q_1= \dots = Q_N=0$,
just as in the computation of the norm of the state
and hence the expectation value of 
 ${\cal O}_{i,i+1}^{\rm d}$ reduces
to an expectation value in the SM model
with Boltzmann weights $W^0$. This is a property
of all diagonal operators which allow
their exact evaluation, provided they are
known in the underlying exactly solved model.

In the case of the operator  
$\sigma^z_{i,a} \sigma^z_{i+1,a}$, 
its correlator is equivalent 
to the SM expectation value 

\begin{equation}
P \equiv  \langle \sigma^z_{i,a} \;  \sigma^z_{i+1,a} \rangle_{\rm ATPA}
= \langle s_{i,a} \; s_{i+1,a} \rangle_{\rm 6-vertex} 
\label{30}
\end{equation}

\noindent where $s= 1, -1$ is related to the link variable 
$\alpha =0, 1$ by the equation $s= 1- 2 \alpha$.  
In appendix B we shall compute this quantity using
the exact solution of the six-vertex model.

Let us next consider an off-diagonal operator 
${\cal O}_{i}^{\rm od}$
acting on the $i^{\rm th}$ leg with 
matrix elements  
$\hat{{\cal O}}_{ {\bf m_i},{\bf m'_i}}^{\rm od}
= \langle {\bf m_i} | {\cal O}_{i}^{\rm od} | {\bf m'_i} \rangle$.
Its expectation value will be  given by,

\begin{equation}
\langle {\cal O}_{i}^{\rm od} \rangle
= {\rm Tr} \left( \rho_i \;   \hat{{\cal O}}_{}^{\rm od} \right)
\label{31}
\end{equation}

\noindent where $\rho_i$ is the density matrix of the $i^{\rm th}$
leg whose entries are,

\begin{align}
\rho_{{\bf m_i}, {\bf m'_i}} &  = 
\frac{1}{\langle \psi|\psi\rangle}
\; \sum_{{\bf m}'s \neq {\bf m_i}\;  {\rm or} \; {\bf m'_i}}
 (T^{\rm row}_{{\bf m_1}, {\bf m_2}})^2  \dots \label{32} \\
& T^{\rm row}_{{\bf m_{i-1}}, {\bf m_i}}
 T^{\rm row}_{{\bf m_{i-1}}, {\bf m'_i}}
 T^{\rm row}_{{\bf m_{i}}, {\bf m_{i+1}}}
 T^{\rm row}_{{\bf m'_{i}}, {\bf m_{i+1}}}
\dots  (T^{\rm row}_{{\bf m_N}, {\bf m_1}})^2 \nonumber 
\end{align}

Using again eq.(\ref{28}) in the thermodynamic
limit we can write $\rho_i$ as

\begin{align}
 \rho_{{\bf m_i}, {\bf m'_i}} &  = 
\frac{1}{\Lambda_{\rm max}^2}
\; \sum_{ {\bf m_{i-1}},  {\bf m_{i+1}}}
v^l_{{\bf m_{i-1}}} 
T^{\rm row}_{{\bf m_{i-1}}, {\bf m_i}}
 T^{\rm row}_{{\bf m_{i-1}}, {\bf m'_i}}
\label{33} \\
& \hspace{3cm} \cdot 
T^{\rm row}_{{\bf m_{i}}, {\bf m_{i+1}}}
 T^{\rm row}_{{\bf m'_{i}}, {\bf m_{i+1}}}
 v^r_{{\bf m_{i+1}}} 
\nonumber 
\end{align}

\noindent where  $v^{l/r}_{{\bf m}}$ are the left 
and right eigenvectors with 
highest eigenvalue $\Lambda_{\rm max}$ of the transfer
matrix $T(W^0)$. Eq.(\ref{33}) shows that the 
regions located above or below of a given 
leg behave as if they were in a single coherent state, 
which in some cases can be identified with 
the ground state of the underlying quantum mechanical model. 
This is indeed the case if we assume 
that $T(W^0)$ is a symmetric matrix, which is achieved
in the six-vertex model 
if  the Boltzmann weights $a$ and $b$ are equal. 
From now on we shall
assume the latter condition which implies
that  $v^{l}_{{\bf m}} =  v^{r}_{{\bf m}} =  v_{{\bf m}}$.

The computation of (\ref{31}) is in general quite
difficult depending on the operator in question.
An approximation can however be made using the
following result. If 
${\cal O}_{i}^{\rm od}$ is a positive definite operator
then 

\begin{equation}
\langle {\cal O}_{i}^{\rm od} \rangle
\leq \langle v|    \hat{{\cal O}}_{}^{\rm od} | v \rangle 
\label{34}
\end{equation}

The proof of (\ref{34}) uses the
Perron-Frobenius theorem and the Schwarz inequality
and it is  similar to the one given in Appendix A to prove
eq.(\ref{23}). For diagonal operators acting on a 
leg, eq.(\ref{34}) becomes an equality.

\section{The six-vertex ATPA}
\label{sec5:level1}

The considerations made above 
suggest that an ATPA based on the six-vertex model
should be a reasonable approximation to the
ground state of the following Hamiltonian:

\begin{align}
& H = H_{\rm leg} + H_{\rm rung}
\label{35} \\
& H_{\rm leg} = 
-  \frac{1}{2} \sum_{i=1}^N \sum_{a=1}^L
\left( \sigma^x_{i,a} \sigma^x_{i,a+1} + 
\sigma^y_{i,a} \sigma^y_{i,a+1} + \Delta_0 
\sigma^z_{i,a} \sigma^z_{i,a+1} \right) 
\nonumber \\
& H_{\rm rung} = \frac{1}{2} J'
 \sum_{i=1}^N \sum_{a=1}^L
\sigma^z_{i,a} \sigma^z_{i+1,a}
\nonumber 
\end{align}

\noindent which is a combination of the XXZ
Hamiltonian along the legs and an Ising one
along the rungs. The latter choice is motivated
by the absence of quantum fluctuations
across the legs.  This model has 
also been studied in reference \cite{jose}
using bosonization techniques.  
Using the Hellman-Feynman 
theorem and eqs.(\ref{30}) and (\ref{34}),
one can find the following lower bound of 
the energy per site of  the ATPA

\begin{equation}
E(\Delta, \Delta_0, J')=
E_0(\Delta) + (\Delta_0 - \Delta) \; \frac{\partial E_0}{\partial \Delta}
+ \frac{1}{2} J'  \; P(\Delta)
\label{36}
\end{equation}

\noindent 
where $\Delta$ is the anisotropy parameter 
associated to the Boltzmann weights $W^0$, i.e.

\begin{align}
& \Delta = \frac{a_0^2 + b_0^2 - c_0^2}{2 a_0 b_0} \label{37} \\
& a_0 = a^2, b_0 = b^2, c_0 = c^2  \nonumber 
\end{align}

\noindent $P(\Delta)$ is the expectation value 
defined in eq.(\ref{30}), and $E_0(\Delta)$ 
is the GS energy per site of the XXZ model
with anisotropy $\Delta$. 

The problem is: fixing
$\Delta_0$ and $J'$, find the value of $\Delta$
that minimizes the total energy (\ref{36}), i.e.

\begin{equation}
\Delta = \Delta(\Delta_0, J')
\label{38}
\end{equation}

\noindent It is easy to see that if $J'=0$ 
then $\Delta = \Delta_0$.

In the antiferromanetic region ($AF$) of the
XXZ model, i.e.  $\Delta < -1$,  the parametrization
of the Boltzmann weights is given by \cite{baxter},

\begin{align}
&  \Delta = - {\rm cosh} \lambda, \;\; \lambda > 0 \nonumber \\
& a_0 = \rho \; {\rm sinh} \frac{\lambda -v}{2} \label{39} \\
& b_0 = \rho \; {\rm sinh} \frac{\lambda + v}{2} \nonumber \\
& c_0 = \rho \;  {\rm sinh}  \lambda  \nonumber 
\end{align}

\noindent where $\rho$ is an overall factor
and $v$ is the spectral parameter that is  set to zero
in order to have a symmetric transfer matrix. 

The total energy and its derivative in the 
$AF$  region 
can be found from the Bethe ansatz solution and they read 
\cite{baxter},

\begin{align}
E_0 &  =   \frac{1}{2} {\rm cosh} \lambda - {\rm sinh} \lambda
- 4 \; {\rm sinh} \lambda \sum_{m=1}^\infty
\frac{1}{{\rm e}^{2 m \lambda} +1}
\label{40} \\
\frac{\partial E_0}{\partial \Delta} &  = 
- \frac{1}{2} + \frac{1}{ {\rm tanh} \lambda} 
+  \frac{4}{ {\rm tanh} \lambda}   \sum_{m=1}^\infty
\frac{1}{{\rm e}^{2 m \lambda} +1}
\label{41} \\
& - 8  \sum_{m=1}^\infty
\frac{m \; {\rm e}^{2 m \lambda}}{({\rm e}^{2 m \lambda} +1)^2}
\nonumber 
\end{align}

The matrix element $P$ in this region is derived in Appendix B
and it reads,

\begin{equation}
 P = -1 - \frac{2}{{\rm sinh}^2(\lambda/2)}
+ \frac{4}{{\rm tanh}(\lambda/2)}
\sum_{m=1}^\infty \frac{{\rm sinh} m \lambda}{ {\rm cosh}^2 m \lambda} 
\label{42} 
\end{equation}

In the critical  region ($C$), i.e. $ -1 < \Delta < 1$, 
  the parametrization
of the Boltzmann weights is given by \cite{baxter},

\begin{align}
&  \Delta = - {\rm cos} \mu, \;\; 0 < \mu < \pi \nonumber \\
& a_0 = \rho \; {\rm sin} \frac{\mu -v}{2} \label{43} \\
& b_0 = \rho \; {\rm sin} \frac{\mu + v}{2} \nonumber \\
& c_0 = \rho \;  {\rm sin}  \mu  \nonumber 
\end{align}

The GS energy per site and its derivative read \cite{baxter}

\begin{align}
E_0 &  = \frac{1}{2} {\rm cos} \mu - \frac{\rm sin \mu}{\mu}
\left[ 2 \; {\rm log} 2 - 2 \pi 
\int_0^\infty dx \; 
\frac{ {\rm log} \; {\rm cosh} \mu x }{ 
{\rm sinh}^2 \pi x } \right]
\label{44} \\
\frac{\partial E_0}{\partial \Delta} & = 
- \frac{1}{2} 
+ \frac{2 \pi}{ \mu} 
\int_0^\infty dx \; 
\frac{ x \; {\rm tanh} \mu x }{ 
{\rm sinh}^2 \pi x } 
+ \left( \frac{1}{\mu \; {\rm tan} \mu}
- \frac{1}{\mu^2} \right) \label{45} \\ 
& \times \left[ - 2 \; {\rm log} 2 +
2 \pi \int_0^\infty dx \; 
\frac{ {\rm log} \; {\rm cosh} \mu x }{ 
{\rm sinh}^2 \pi x }  \right] & \nonumber 
\end{align}

\noindent and $P$ is given by ( see Appendix B),

\begin{equation}
P = 1 - \frac{4 \pi}{\mu \; {\rm tan}(\mu/2)}
\int_0^\infty \; dx \;  \frac{ {\rm sinh}^2 \mu x}
{{\rm cosh} \mu x \;  {\rm sinh}^2 \pi x}
\label{46}
\end{equation}

\begin{figure}[t]
\includegraphics[width= 7 cm]{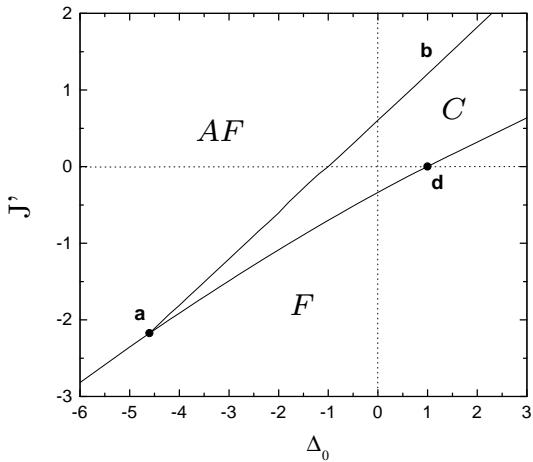}
\caption{
Phase diagram of the Hamiltonian (\ref{35}) 
obtained using the six-vertex ATPA.  
The point $a$ corresponds to $\Delta_0 = -4.6$
and $J'= -2.17$, $b$ is a generic
point on the line $AF/C$ above the point
$a$,  and finally $d$ corresponds to  $\Delta_0 = 1$
and $J'=0$. 
}
\label{fig1}
\end{figure}

Figure 1 shows the phase diagram of the
Hamiltonian (\ref{35}) obtained
by minimization of the energy 
(\ref{36}). 
We recall that (\ref{36}) is 
a lower bound of the energy of the ATPA,
and hence does not yield  an upper
bound of the exact GS energy of
(\ref{35}). 

The region denoted $AF$ in fig. 1 
corresponds to the cases
where $ \Delta(\Delta_0, J')$ lies inside
the antiferromagnetic 
regime $ \Delta < -1$. 
The region $C$ describes the critical
regime, i.e. 
$- 1 < \Delta < 1$, while the region $F$ 
denotes the cases where $\Delta =1$. 
The phase boundaries between these regions 
have different properties. 
The $AF/C$ boundary 
line ${\bf ab}$ corresponds to $\Delta = -1$,  
and hence the transition between the $AF$
and $C$ phases seems to be continuous. 
Below the point ${\bf a}$ 
the value of  $\Delta$, near the line
$AF/F$ but on the $AF$ side, is smaller
than $-1$, indicating  that the
$AF/F$ boundary is discontinuous.  
The $C/F$ line ${\bf ad}$ is also discontinuous, 
meaning that $\Delta$ jumps across it.
Finally, there is no discontinuities across the
$C/F$ boundary above the point ${\bf d}$.  

In reference \cite{jose} the Hamiltonian
(\ref{35}) was studied using bosonization,
mean field and  renormalization group (RG) 
methods. Disregarding the inter-leg forward
scattering and umklapp terms, that arise
upon bosonization, the main conclusion
of reference \cite{jose}  is the existence
of an AF region whenever $|J'| > 2 \Delta_0$ 
(mean field result) or  $|J'| > 4 \Delta_0$
( RG result). Furthermore,  
the inter-chain forward scattering terms 
can be taken into account \cite{sierra} using the sliding Luttinger
liquid approach of references \cite{emery00,carpentier}.
In \cite{sierra} it was 
shown that the  effect of the inter-chain forward
scattering is to modify
the phase boundaries separating the AF and C regions
in an asymmetric way. Indeed the AF region
appears when $J' > C_+  \Delta_0$ 
( if $J' >0$) and $- J' >  C_- \Delta_0$ 
( if $J' < 0$), with $C_+ \neq C_-$. 
Hence the results of references \cite{jose,sierra},
which should be valid in the weak coupling regime
$|\Delta_0| , |J'| << 1$, 
suggest  that the system should be in an AF phase
whenever the legs are antiferromagnetic, i.e. 
$\Delta_0 < 0$.  This is in contradiction with
the ATPA result where there exist critical region  
with  $\Delta_0 < 0$. On the other hand
the ATPA agrees with the aforementioned works 
on  the existence of large regions
in the phase diagram  where the system is critical, 
which  we identify with the sliding or smectic Luttinger
liquid fixed points of \cite{emery00,carpentier}.

\section{The eight-vertex ATPA}
\label{sec6:level1}

An ATPA closely related to the
six-vertex one can be built from 
the Baxter's eight-vertex
model,  whose Boltzmann weights are
those of the six-vertex model plus 
two new weights $W^{1 1}_{0 0} = W^{0 0}_{1 1}=d$ \cite{baxter}.
The conservation law (\ref{18}) now becomes,

\begin{equation}
W_{\alpha, \xi}^{\beta,  \eta} = 0
\;\; {\rm unless} \;\; \alpha + \xi = \beta + \eta, \; \;  ({\rm mod} \; 2) 
\label{47}
\end{equation}

The transfer
matrix ${\cal T}^{\rm col}$ also breaks 
into block matrices $T^{\bf Q}$ with the 
difference   that $Q_i$ only takes two values 0 and 1, 
since  $Q_i = -1 = 1 (\rm mod \; 2)$. 

The eight-vertex ATPA can be taken as the 
an ansazt for the GS of the following Hamiltonian:

\begin{align}
& H = H_{\rm leg} + H_{\rm rung}
\label{48} \\
& H_{\rm leg} = 
-  \frac{1}{2} \sum_{i,a}
\left(J_x \sigma^x_{i,a} \sigma^x_{i,a+1} + 
J_y \sigma^y_{i,a} \sigma^y_{i,a+1} + J_z
\sigma^z_{i,a} \sigma^z_{i,a+1} \right) 
\nonumber \\
& H_{\rm rung} = \frac{1}{2} J'
 \sum_{i,a }
\sigma^z_{i,a} \sigma^z_{i+1,a}
\nonumber 
\end{align}

The phase diagram of this model can be 
worked out using the Baxter's exact solution 
of the eight-vertex model,
as we did for the
six-vertex one in the previous section. 
The results will be presented elsewhere.

\section{The ATPA and stripes}
\label{sec7:level1}

An interesting feature of the six 
and the eight-vertex ATPA's is their possible
connection with the stripes in high $T_c$
superconductors, specially when regarded
as electronic liquid crystals \cite{kivelson98,emery00,carpentier}. 
In this section we shall briefly explore this issue
which deserves a more detailed study in the future.

The first observation is that the eight-vertex
Hamiltonian (\ref{48}) ( and similarly the six-vertex one
(\ref{35}) )  can be Jordan-Wigner 
transformed 
onto the following spinless fermion Hamiltonian, 

\begin{equation}
\begin{aligned}
 H  = - \frac{1}{2}  \sum_{i,a} &  \left[ 
\;\;  (J_x + J_y) (\psi^\dagger_{i,a} 
\psi_{i+1,a} + h.c.) \right.  \\
&  +  (J_y - J_x) (\psi_{i,a} 
\psi_{i+1,a} + h.c.) \\
& + 2 J_z (n_{i,a} - \frac{1}{2})
( n_{i+1,a} - \frac{1}{2} ) \\
& \left.  - 2 J'  (n_{i,a} - \frac{1}{2})( n_{i,a+1} - \frac{1}{2}) \right]
\end{aligned}
\label{49}
\end{equation}

\noindent 
which  describes the motion of holons 
along the legs of a 2D lattice ( term $(J_x + J_y)$), 
which are coupled by 
density-density interactions ( term $J'$),  together with 
pair tunneling between the legs and the environment  
( term $(J_x - J_y)$) \cite{emery97}. Upon bosonization
eq.(\ref{49}) has a structure similar, 
but not identical, to the ``smectic'' Hamiltonian in the
spin gap case considered in reference \cite{emery00}
and the spinless sliding Luttinger model of reference
\cite{carpentier}. 
Indeed, the smectic symmetry $\phi_ a \rightarrow 
\phi_a + \alpha_a$ \cite{emery00}, where $\phi_a$ is the boson
field of the $a^{\rm th}$ leg, is the dual 
version of the standard $U(1)$ symmetry of the 
six-vertex model,  which corresponds to $\theta_a \rightarrow
\theta_a + \alpha_a$, where $\theta_a$ is the dual boson 
\cite{affleck}. The CDW coupling among the stripes in
\cite{emery00,carpentier} corresponds to the term $J'$, while
the Josephson tunneling is somehow reflected by the
pair creation and annihilation terms. 
Assuming these correspondences it is quite 
natural to conjecture a relationship between the smectic
phases of references \cite{emery00,carpentier} 
and the corresponding phases
of the eight-vertex model. The stripe crystal phase 
should correspond to the antiferromagnetic phase, 
the smectic superconducting phase should correspond
to the disordered phase and finally, the smectic metal
should be associated to the critical phase, which 
is the one of the six-vertex model when $ -1 < \Delta < 1$.

\section{Conclusions}
\label{sec8:level1}

In this paper we have proposed a new class of Anisotropic
Tensor Product Ansatzs (ATPA) using the Boltzmann
weights of classical Statistical Mechanics vertex models. 

We have shown that the computation of the norm and some
observables simplifies enormously,  becoming exact whenever
the underlying SM model is exactly solvable. 

The strong anisotropy of the ATPA's is reflected in the
absence of quantum fluctuations across the legs of the
2D lattice, a property which suggests a possible
connection with some current models of stripes. 

We have studied the ATPA based on the six-vertex model,
as a trial state for the ground state of a Hamiltonian
given by the sum of XXZ Hamiltonians along the legs 
of a 2D lattice, which are coupled by an Ising term. 
Using the exact solution of the six-vertex model 
we have proposed the phase diagram of this model
and compared it with the one obtain with other methods
\cite{jose,sierra}.

We have suggested  a connection 
between the six-vertex and eight-vertex ATPA's, 
and their associated 2D Hamiltonians, with 
the smectic stripe phases considered in references
\cite{emery00,carpentier}. 

Let us finally comment on 
the relation between the ATPA and the DMRG. 
As we explained in section IV, the link variables
along the legs and the rungs 
of the SM vertex model can be of different type.
For example we can choose the rung variables
$\xi, \eta$ to take only two values, say 
0 and 1,  as in the six vertex model,  while the legs variables
$\alpha, \beta $ can take a large 
number of values, say $1, 2, \dots, m$, 
as in the DMRG. The ATPA so constructed would 
have  a spin 1/2 at each site  with strong 
correlations along the legs. This state would be  
a sort of anisotropic DMRG state with a stripe like
structure built in. The problem is to device
an algorithm to update the local
weights.

\subsection*{Acknowledgments}

We would like to thank T. Nishino
for conversations. 
This work has been partially supported by the Spanish grant PB98-0685.

\section*{Appendix A: Highest eigenvalue of $\T$}
\label{secA:level1}

In this appendix we shall give a proof of eq.(\ref{23})
under the condition that all the TPA amplitudes
are non negative.  
Let us call $\Lambda^{\bf Q}_0$ the largest 
eigenvalue of the transfer matrix $\T$ defined
in eq.(\ref{22}). 
The statement is that

\begin{equation}
\Lambda^{{\bf Q}}_0 \leq  
\Lambda^{{\bf Q}= 0}_0  , \;\; \forall {\bf Q}
\label{a1}
\end{equation}

\noindent 
Choosing  two  
vectors $\chi$ and $\phi$,  with
positive entries and scalar product equal to 1,

\begin{align}
& \langle \chi|\phi \rangle=1 \label{a2} \\ 
& \chi_{\ba} = u_{\ba}^2,\;\;  \phi_{\ba}= v_{\ba}^2, 
\;\;  u_{\ba},  v_{\ba} > 0
\nonumber 
\end{align}

\noindent 
one has by the definition of  $\Lambda^{\bf Q}_0$,

\begin{equation}
\langle \chi| \T |\phi \rangle \leq   \Lambda^{\bf Q}_0
\label{a3} 
\end{equation}

\noindent 
where the equality holds whenever $\chi$ and
$\phi$ are the left and right eigenvectors of $\T$
respectively (recall that, by  the Perron-Frobenius theorem,
the eigenvector of $\T$, with highest eigenvalue, 
has all its entries positive). 

\noindent 
Using (\ref{9}) and (\ref{22})
we can write  the LHS of (\ref{a3}) as

\begin{align}
\langle \chi| \T |\phi \rangle & =   \sum_{I} x_I^{\0} x_I^{\bf Q}
 \label{a4} \\
& = \sum_{\ba, \bb, {\bf m}}  u_{\ba}^2  v_{\bb}^2
A^{\rm col}_{\ba,\bb}[{\bf m}] 
A^{\rm col}_{\ba + {\bf Q},\bb + {\bf Q}}[{\bf m}] 
\nonumber 
\end{align}

\noindent where $I$ denotes the triple
$(\ba,\bb, {\bf m})$ 
and $x_I^{\bf Q}$ stands for

\begin{align}
& x_I^{\bf Q} =  u_{\ba}  v_{\bb}
A^{\rm col}_{\ba + {\bf Q},\bb + {\bf Q}}[{\bf m}] 
\label{a5} 
\end{align}

\noindent it turns out that

\begin{equation}
\sum_I \left( x_I^{\bf Q} \right)^2 \leq \Lambda_0^{\0}, \;\;
\forall {\bf Q}
\label{a6}
\end{equation}

\noindent 
On the other hand, the Schwarz inequality

\begin{equation}
|  \sum_{I} x_I^{\0} x_I^{\bf Q}| 
\leq \sqrt{
\left(  \sum_{I} x_I^{\0} x_I^{\0} \right)
\left(  \sum_{J} x_J^{\bf Q} x_J^{\bf Q} \right) }
\label{a7}
\end{equation}

\noindent implies

\begin{equation}
|\langle \chi| \T |\phi \rangle| \leq   \Lambda^{\0}_0
\label{a8} 
\end{equation}

\noindent Hence, choosing $\chi$ and $\phi$ the left
and right eigenvectors of $\T$ one derives
the desired result (\ref{a1}).

\section*{Appendix B: The two point correlator $P(\Delta)$  }
\label{secB:level1}

In this appendix we indicate how to
compute the expectation value

\begin{equation}
P = \langle s_{i,a} \; s_{i+1,a} \rangle_{\rm 6-vertex} 
\label{b1}
\end{equation}

\noindent in the six-vertex model with Boltzmann weights
$a,b$ and $c$ when $a=b$, which is the case
under study. This quantity is similar, but not identical,
to the polarizability $P_0 = \langle \alpha_1 \rangle$
defined by Baxter \cite{baxter}. 

The partition function of the six-vertex model can be
expanded as

\begin{equation}
Z = \sum a^{n_1 + n_2} \; b^{n_3 + n_4} c^{n_5 + n_6}
\label{b2}
\end{equation}

\noindent where $n_1$ and $n_2$ are the 
number of vertices with Boltzmann weight $a$, etc. 
The weights $a$ and $b$ contribute to $P$ with
$+1$ while $c$ does it  with $-1$,
hence $P$ is given by the formula,

\begin{align}
& P = \lim_{N,L \rightarrow \infty}
\frac{1}{N L} \frac{1}{Z}
\left( a \frac{\partial}{ \partial a} 
+  b \frac{\partial}{ \partial b} 
-  c \frac{\partial}{ \partial c} \right) Z 
\label{b3} \\
& = - \left( a \frac{\partial}{ \partial a} 
+  b \frac{\partial}{ \partial b} 
-  c \frac{\partial}{ \partial c} \right) f
\nonumber
\end{align}

\noindent where $f$ is the free energy per site
in the units $k_B T = 1$.  
It is important to realize that the derivatives
in (\ref{b3})
are performed keeping the remaining ones unchanged. 
Eq.(\ref{b3}) assumes that $a,b$ and $c$ 
are independent quantities, however if 
$a=b$ the formula for $P$ becomes,

\begin{align}
& P_{a=b}
= - \left( a \frac{\partial}{ \partial a} 
-  c \frac{\partial}{ \partial c} \right) f
\label{b4}
\end{align}

In the $AF$ region the free energy $f$ is given,
in the parametrization (\ref{39}), by \cite{baxter},

\begin{align}
& - f =  {\rm log}\; a + \frac{ \lambda + v}{2} 
+ \sum_{m=1}^\infty \frac{ {\rm e}^{-m \lambda}\; {\rm sinh}\; m(\lambda +
v)}{ m \;  {\rm cosh} \; m \lambda} \label{b5}
\end{align}

\noindent while in the critical region one has, 
in the parametrization (\ref{43}) \cite{baxter}

\begin{align}
& - f =  {\rm log}\; a 
+ \int_{- \infty}^\infty \frac{dx}{x} 
 \frac{  {\rm sinh}\; (\lambda + v)x 
\; {\rm sinh} \; (\pi - \mu) x }{
{\rm sinh} \; \pi x \; {\rm cosh} \mu x}
\label{b6}
\end{align}

\noindent 
Using eqs.(\ref{b4}), (\ref{b5}) , (\ref{b6}),
(\ref{39}) and  (\ref{43}), with $v$ set equal to zero, 
one can derive eqs.(\ref{42}) and (\ref{46})
yielding $P(\Delta)$.

In figure 2 we plot $P(\Delta)$ in the $AF$ and $C$ regions. 
In the $AF$ region one has 
$ -1 < P < -1/3$, while in the $C$
region $ -1/3 < P < 1$. At the isotropic point
$\Delta = -1$ one finds $P= -1/3$, while in the
XY model, i.e.  $\Delta=0$, the result is  $P=0$.  
In all the ferromagnetic ($F$)  region, i.e. $ \Delta > 1$, 
one has $P=1$.

\begin{figure}[t]
\includegraphics[width= 7 cm]{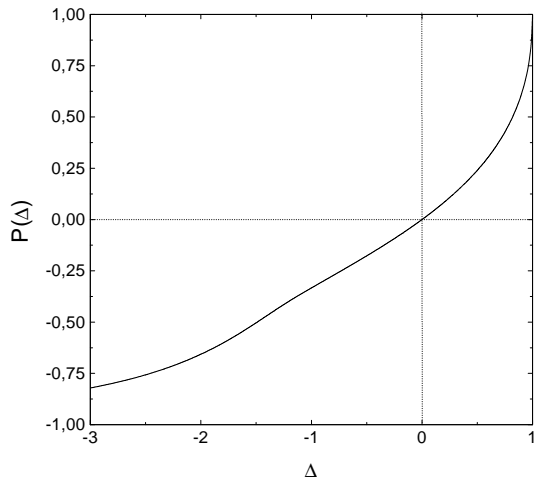}
\caption{Plot of $P(\Delta) = 
\langle s_{i,a} \; s_{i+1,a} \rangle_{\rm 6-vertex}$ 
in the AF region (eq.(\ref{42})) and critical region (eq.(\ref{46})). 
}
\label{fig2}
\end{figure}

\end{document}